\documentstyle{article}

\title{Sigma-models having supermanifolds as target spaces.}
\author{Albert Schwarz\thanks {Research supported in part by NSF grant No.
DMS-9201366}\\
Department of Mathematics, University of California,\\ Davis, CA 95616\\
ASSCHWARZ@UCDAVIS.EDU}
\date{}

\begin{document}
 \maketitle
 \smallskip
 \begin{abstract}

    We study a topological sigma-model ($A$-model) in the case when the
target space is an ($m_0|m_1$)-dimensional supermanifold. We prove  under
certain conditions that such a model is equivalent to an $A$-model having
an ($m_0-m_1$)-dimensional manifold as a target space. We use this result
to prove that in the case when the target space of  $A$-model is a
complete intersection in a toric manifold, this $A$-model is equivalent
to an $A$-model having a toric supermanifold as a target space.

 \end{abstract}

    Our goal is to study a two-dimensional topological  $\sigma$-model
($A$-model). Sigma-models having supermanifolds as target spaces were
considered in an interesting paper [5]. However, the approach of [5]
leads to a conclusion that in the case when the target space of $A$-model
is a supermanifold the contribution of rational curves to correlation
functions vanishes (i.e. these functions are essentially trivial). In our
approach $A$-model having a $(m_0|m_1)$-dimensional supermanifold as a
target space is not trivial, but it is equivalent to an  $A$-model with
$(m_0-m_1)$-dimensional target space. We hope, that this equivalence can
be used to understand better the mirror symmetry, because it permits us
to replace most interesting target spaces with supermanifolds having
non-trivial Killing vectors and to use  $T$-duality.

     We start   with a definition of $A$-model given in [1]. This
definition can be applied to the case when the  target space is a complex
K\"ahler supermanifold $M$. Repeating the consideration of [1] we see
that the correlation functions can be expressed in terms of rational
curves in $M$, i.e. holomorphic maps of $CP^1$ into $M$. (We restrict
ourselves to the genus $0$ case and assume that the situation is generic;
these restrictions will be lifted in a forthcoming paper [8]).

   Let us consider for simplicity the case when ($m_0|m_1$)-dimensional
complex  supermanifold $M$ corresponds to an   $m_1$-dimensional
holomorphic
vector bundle $\alpha$ over an  $m_0$-dimensional complex manifold $M_0$
(i.e. $M$ can be obtained from the total space of the bundle $\alpha$ by
means of reversion of parity of the fibres.) The natural map of $M$ onto
$M_0$ will be denoted by $\pi$. To construct the correlation functions of
the $A$-model with the target space $M$ we should fix real  submanifolds
$N_1,...,N_k$ of $M_0$ and the points $x_1,...,x_k\in CP^1$. For every
two-dimensional homology class $\lambda \in H_2(M,{\bf Z})$ we consider
the space $D_{\lambda}$ of  holomorphic maps $\varphi$ of $CP^1$ into $M$
that transform $CP^1$ into a cycle $\varphi (CP^1)\in \lambda$ and
satisfy the conditions $\pi (\varphi (x_1))\in N_1,...,\pi(\varphi
(x_k))\in N_k$. (We identify the  homology of $M$ with the  homology of
its body $M_0$; the condition $\varphi (CP^1)\in \lambda$ means that the
image of the fundamental  homology class of $CP^1$ by the homomorphism
$(\pi \varphi )_*:H_2(CP^1,{\bf Z})\rightarrow H_2(M_0,{\bf Z})$ is equal
to $\lambda$.) The space $D_{\lambda}$ contributes to the correlation
function under consideration only if
\begin {equation}
2m_0-(q_1+...+q_k)+<c_1(T),\lambda >=2m_1(<c_1(\alpha ),\lambda >+1)
\end {equation}
where $c_1(T)$ is the first Chern class of the tangent bundle to $M,
c_1(\alpha)$ is the first Chern class of the bundle $\alpha$ and
$q_i=2m_0-\dim N_i$ denotes the codimension of $N_i$. If $\varphi \in
D_{\lambda}$ then  $\pi (\varphi) \in D_{\lambda}^0$ where  $
D_{\lambda}$ is the space of holomorphic maps $\phi :CP^1\rightarrow M_0$
obeying $\phi (CP^1) \in \lambda$ and $\phi (x_1)\in N_1,...,\phi
(x_k)\in N_k$.

  Let us consider a holomorphic vector bundle $\xi _{\lambda}$ over
$D^0_{\lambda}$ having the vector  space of holomorphic sections of the
pullback of $\alpha$ by the map $\phi \in D^0_{\lambda}$ as a fiber over
$\phi$. It is easy to check that $D_{\lambda}$ can be obtained from the
total space of $\xi _{\lambda}$ by means of parity reversion in the
fibers. It follows from the index theorem that the virtual dimension of
$D^0_{\lambda}$ is equal to $d_1=2m_0-\sum q_i+2<c_1(T),\lambda >$; our
assumption that the situation is generic means that $d_1=\dim
D^0_{\lambda}$. The Riemann-Roch theorem together with equation (1)
permits us to say that the dimension of the fiber of $\xi _{\lambda}$ is
equal to $d_2=2m_1(<c_2(\alpha),\lambda >+1)$ and coincides with $d_1$.
We see that the even dimension $d_1$ of $ D_{\lambda}$ coincides with its
odd dimension $d_2$. The contribution of $D_{\lambda}$ into the
correlation function can be expressed in terms of the Euler number of the
vector bundle $\xi _{\lambda}$ (see [2] or [3] for explanation of similar
statements in a little bit different situations).

  Let us consider now a holomorphic section $F$ of $\alpha$. We will
assume that the zero locus of $F$ is a manifold and denote this manifold
by $X$. The K\"ahler metric on $M$ induces a K\"ahler metric on $X$;
therefore we can consider an $A$-model with the target space $X$ . We'll
check that the  correlation functions of this $A$-model coincide with
the  correlation functions of the $A$-model with target space $M$. More
precisely, the  correlation function of $A$-model with target space $M$
constructed by means of submanifolds $N_1,...,N_k \subset M_0$ coincides
with the  correlation function of  $A$-model with target space $X$
constructed by means of submanifolds $N_1^{\prime}=N_1\cap
X,...,N_k^{\prime}=N_k\cap X$ of the manifold $X$. (Without loss of
generality we can assume that $N_i^{\prime}=N_i\cap X$ is a submanifold).
To prove this statement we notice that using the section $F$ of $\alpha$
we can construct a section $f_{\lambda}$ of $\xi _{\lambda}$ assigning to
every map $\phi \in D_{\lambda}^0$ an element $f_{\lambda}(\phi)=F\cdot
\phi$ of the fiber of $\xi_{\lambda}$ over  $\phi \in D_{\lambda}^0$. It
is easy to check that zeros of the section $f_{\lambda}$ can be
identified with holomorphic maps  $\phi \in D_{\lambda}^0$ satisfying
$\phi (CP^1)\subset X$. The number of such maps enters the expression
for  correlation functions of $A$-model with target space $X$. From the
other side this number coincides with the Euler number of $\xi
_{\lambda}$ entering corresponding expression in the case of target space
$M$. This remark proves the coincidence of  correlation functions for the
target space $M$ with  correlation functions for the target space $X$.
Let us stress, however, that not all  correlation functions for the
target space $X$ can be obtained by means of above construction. Using
the language of cohomology one can say that a  correlation function of an
$A$-model  with the target space $X$ corresponds to a set of cohomology
classes $\nu _1,...,\nu _k\in H(X,{\bf C})$.
  Such a  correlation function is equal to a  correlation function of an
$A$-model with the target space $M$ if there exist cohomology classes
$\tilde {\nu }_1,...,\tilde {\nu} _k\in H(M_0,{\bf C})$ obeying
$\nu_1=i^*\tilde {\nu}_1,...,\nu_k=i^*\tilde {\nu}_k$. (Here $i$ denotes
the embedding of $X$ into $M_0$. We used the fact that cohomology class
$\nu _i\in H(X,{\bf C})$, dual to $N_i^{\prime}=N_i\cap X$, is equal to
$i^*\tilde {\nu}_i$ where $\tilde {\nu}_i\in H(M_0,{\bf C})$ is dual to
$N_i$).

To prove that correlation functions of $A$-model having a supermanifold
as a target space coincide with correlation functions of an ordinary
$A$-model we used arguments similar to the arguments, utilized in [4].
One can say that our consideration reveals the geometric meaning of
Kontsevich's calculation.

  We define a Calabi-Yau supermanifold (CY- supermanifold) as a
K\"ahler supermanifold having trivial canonical line bundle. (Recall that
the fiber $K_x$ of canonical line bundle $K$ over a complex supermanifold
$M$ can be defined as $m(T_x(M))$ where $T_x(M)$ denotes the tangent
space at the point $x\in M$ and $m(E)$
denotes the one-dimensional linear space of complex measures on a complex
linear space $E$.) It is easy to prove that the canonical  bundle over
$X$ in the construction above can be obtained by means of restriction of
the canonical bundle over $M$; therefore if $M$ is a CY-supermanifold,
$X$ is a CY-manifold. The proof is based on the following simple remark.
If $E=E_0+E_1$ is a linear superspace with even part $E_0$ and odd part
$E_1$ and $A:E_0\rightarrow E_1$ is a surjective linear operator, then
$m(\ker A)$ is canonically isomorphic to $m(E)$. (This follows from the
canonical isomorphisms $m(E)=m(E_0)\otimes m(E_1)^*,\
m(E_1)=m(E_0)\otimes m(\ker A)^*$).

    Let us consider an example when $M_0=CP^n$. For every $k\in{\bf Z}$
we construct a line bundle $\alpha _k$ over $CP^n$ in the following way. We
define the total space $E_k$ of the  bundle $\alpha _k$ taking quotient
of ${\bf C}^{n+2}\setminus \{ 0\} $ with respect to equivalence relation
  \begin {equation}
  (z_1,...,z_{n+2})\sim (\lambda z_1,...,\lambda z_{n+1}, \lambda ^kz_{n+2}).
  \end {equation}
  The projection map $E_k\rightarrow CP^n$ is induced by the map
$(z_1,...,z_{n+1},z_{n+2})\rightarrow (z_1,...,z_{n+1})$. The
$(n|1)$-dimensional complex supermanifold $M_k$ corresponding to the
line bundle $\alpha _k$, can be obtained from superspace $({\bf
C}^{n+1}\setminus \{ 0\} )\times {\bf C}^{0|1}$
by means of identification
\begin {equation}
  (z_1,...,z_{n+1},\theta)\sim (\lambda z_1,...,\lambda z_{n+1}, \lambda
^k\theta),\  \   \   \lambda \in {\bf C}.
\end {equation}
(here $z_1,...,z_{n+1}$ are even coordinates is ${\bf C}^{n+1|1},\theta$
is an odd coordinate). To give another description of the supermanifold
$M_k$ we consider a hypersurface $N_k\subset {\bf C}^{n+1|1}$, determined
by the equation
 \begin {equation}
  \bar {z}_1z_1+...+\bar {z}_{n+1}z_{n+1}+k\bar {\theta}\theta=c,\  \  \
c>0.
\end {equation}
  The restriction of standard symplectic form $\omega$ on ${\bf
C}^{n+1|1}$ to the hypersurface $N_k$ is a degenerate closed $2$-form.
One can factorize $N_k$ with respect to null-vectors of this $2$-form; it
is easy to check that this factorization leads to identification of
points of $N_k$ given by the formula (3) with $|\lambda|=1$. This means
that a
manifold obtained from $N_k$ by means of factorization with respect to
null-vectors of $\omega$ restricted to $N_k$ can be identified with
$M_k$. This construction equippes $M_k$ with symplectic structure
(depending on the choice of $c>0$). Taking into account that $M_k$ is
equipped with a complex structure we can construct a family of K\"ahler
metrics on $M_k$.

   Every homogeneous polynomial $P(z_1,...,z_{n+1})$ of degree $k$
determines a section of the line bundle $\alpha _k$. We see therefore
that $A$-model with the target space $M_k$ is equivalent to $A$-model
with the target space $X_k$ where $X_k$ is the hypersurface in $CP^n$
singled out by the equation $P(z_1,...,z_{n+1})=0$. (Of course, we should
assume that this hypersurface is smooth.) For example in the case $n=4,
\  k=5$, the  $A$-model with target space $M_5$ is equivalent to the
$A$-model on the quintic in $CP^4$. Notice, that in this case $M_5$ is a
CY- supermanifold (more generally, $M_k$ is a CY-supermanifold if $k=n+1$).

   It is important to emphasize that the manifold $M_k$ has $n+1$
commuting holomorphic Killing vectors (i.e. $n+1$-dimensional torus
$T^{n+1}$ acts on $M_k$; corresponding transformations of $M_k$ are
holomorphic and preserve the metric). This means that we can apply the
machinery of $T$-duality to the $\sigma$-model with target space $M_k$;
one can conjecture that $T$-duality is related to mirror symmetry in this
situation. To describe the above mentioned action of $T^{n+1}$ on $M_k$
we represent a point of $T^{n+1}$ as a row $\sigma =(\sigma
_1,...,\sigma_{n+1},\sigma _{n+2})$ where $|\sigma_i|=1,\sigma_1...\sigma
_{n+2}=1$. Every point $\sigma \in T^{n+1}$ determines a transformation
of $M_k$, sending $(z_1,...,z_{n+1},\theta)$ into $(\sigma
_1z_1,...,\sigma _{n+1}z_{n+1},\sigma _{n+2}\theta)$. It is easy to check
that this transformation is holomorphic and isometric.

     One can obtain an essential generalization of the above construction
using the notion of toric supermanifold.

     Let us consider an $(m|n)$-dimensional complex linear superspace
${\bf C}^{m|n}$ with the standard K\"ahler metric $ds^2=\sum _{a=1}^{m+n}
d\bar {z}_a\cdot dz_a$. (We denote coordinates in ${\bf C}^{m|n}$ by
$z_1,...,z_{m+n}$. The  coordinates  $z_1,...,z_m$ are even, the
coordinates $z_{m+1},...,z_{m+n}$ are odd.) Poisson brackets of functions
$\varphi _1=\bar {z}_1z_1,...,\varphi _{m+n}=\bar {z}_{m+n}z_{m+n}$
vanish; therefore corresponding hamiltonian vector fields generate an
action of  $(m+n)$-dimensional torus $T^{m+n}$ on ${\bf C}^{m|n}$. Let us
consider functions $\psi _i=\sum _{k=1}^{m+n} a_{ik}\varphi _k,\  1\leq
i\leq s,\  a_{ik}\in {\bf Z}$.  Corresponding  hamiltonian vector fields
generate an action of  $s$-dimensional torus $T^s\subset T^{m+n}$ on
${\bf C}^{m+n}$. Let us consider a subset $R_{a,c}$ of ${\bf C}^{m|n}$
determined by the equations $\psi _1=c_1,...,\psi_s=c_s$ where
$a=(a_{ij})$ is an arbitrary integer matrix and $c_1,...,c_s$ are
arbitrary numbers. It is easy to see that $R_{a,c}$ is invariant with
respect to the action of $T^{m+n}$, and therefore  with respect to the
action of $T^s\subset T^{m+n}$. We define a supertoric variety $V_{a,c}$
as a quotient of $R_{a,c}$  with respect to this action of $T^s$.

     One can say also that $V_{a,c}$ is obtained from $R_{a,c}$ by means
of factorization  with respect to null-vectors of a degenerate closed
$2$-form on $R_{a,c}$ (of the restriction of standard symplectic form
$\omega$ on ${\bf C}^{m|n}$ to $R_{a,c}$). We will  consider the case
when $V_{a,c}$ is a supermanifold; then this manifold has a natural
symplectic structure (the degenerate form on $R_{a,c}$ determines a
non-degenerate closed $2$-form on $V_{a,c}$.) The manifold $V_{a,c}$ has
also natural complex structure; the complex and symplectic structure
determine together a K\"ahler metric on a toric supermanifold $V_{a,c}$.
To introduce a complex structure on $V_{a,c}$ we notice that $V_{a,c}$
can be represented as quotient $\tilde {R}_{a,c}/\tilde {T}^s$, where
$\tilde {T}^s$ denotes the complexification of the torus $T^s$ and
$\tilde {R}_{a,c}$ denotes the minimal domain in ${\bf C}^{m|n}$, that
contains  $R_{a,c}$ and is invariant with respect to the action of
complexification $\tilde {T}^{m+n}$ of the torus $T^{m+n}$ (The map
corresponding to a point $(\alpha_1,...,\alpha _{m+n})\in \tilde
{T}^{m+n}=({\bf C}\setminus \{ 0\} )^{m+n}$ transforms
$(z_1,...,z_{m+n})$ into $(\alpha_1z_1,...,\alpha _{m+n}z_{m+n})$.) We
mentioned already that the torus $T^{m+n}$ transforms  $R_{a,c}$ into
itself. Taking into account that the action of $T^{m+n}$ preserves
complex and symplectic structure on ${\bf C}^{m|n}$ we arrive at the
conclusion that the action of $T^{m+n}$ on $R_{a,c}$ descends to a
holomorphic action of $T^{m+n}$ on  $V_{a,c}$, preserving K\"ahler
metric. (More precisely, we have an action of $T^{m+n}/T^s$ on $V_{a,c}$,
because $T^s$ acts trivially.)

     If the above consideration is applied to the case when $n=0$ (i.e.
when instead of complex superspace ${\bf C}^{m|n}$ we consider an
ordinary complex linear space ${\bf C}^m$), we obtain the standard
symplectic construction of toric varieties [6],[7]. Let us consider now
an (ordinary) toric manifold $V_{a,c}$ and a holomorphic line bundle
$\alpha$ over $V_{a,c}$. Then we can construct a complex supermanifold
$W_{\alpha}$ corresponding to the line bundle $\alpha$ (supermanifold
obtained from total space of $\alpha$ by means of reversion of parity of
fibers). We will prove that $W_{\alpha}$ can be considered as a toric
supermanifold. The proof uses the well known fact that every line bundle
over toric manifold $V_{a,c}$ is equivariant with respect to the action
of $T^m$ on $V_{a,c}$ [7]. More precisely every one-dimensional
representation of $T^s$ determines an action of $\tilde  {T}^s$ on ${\bf
C}$; such an action is characterized by integers $\kappa _1,...,\kappa _s$. (An
element $\sigma =(\sigma
_1,...,\sigma _s)\in \tilde{T}^s=({\bf C}\setminus \{ 0\} )^s$ generates
a transformation $c\rightarrow \sigma _1 ^{\kappa _1}...\sigma_s ^{\kappa
_s}c$.) Combining this action with action of $\tilde {T}^s$ on $\tilde
{R}_{a,c}$ we obtain an action of $\tilde {T}^s$ on $R_{a,c}\times {\bf
C}$. One can get a complex  line bundle over $V_{a,c}$ factorizing the
projection $\tilde {R}_{a,c}\times {\bf C}\rightarrow \tilde {R}_{a,c}$ with
respect to
the action of $\tilde {T}^s$. One can prove that an arbitrary holomorphic
line bundle over $V_{a,c}$ can be obtained this way. Replacing ${\bf C}$
with ${\bf C}^{0|1}$ in this construction we can describe the
supermanifold $W_{\alpha}$ as a quotient of $\tilde {R}_{a,c}\times {\bf
C}^{0|1}$ with respect to an action of $\tilde {T}^s$. Using this
description we can identify $W_{\alpha}$ with the toric  supermanifold
$V_{\hat {a},c}$. Here $\hat {a}$ denotes the matrix $a$ with additional
column $(\kappa _1,...,\kappa _s)$, where $\kappa _i$ are the integers
characterizing the action of $\tilde {T}^s$ on ${\bf C}$. (In other
words, if $V_{a,c}$ is obtained as a quotient  $R_{a,c}/T^s$ where
$R_{a,c}$ is defined by the equations $\psi_1=c_1,...,\psi _s=c_s$
in ${\bf C}^m$ then $V_{\hat{a},c}$ is a quotient
$R_{\hat{a},c}/T^s$, where $R_{\hat{a},c}$ is determined by the
equations
$\psi_1+\kappa _1\bar {\theta}\theta =c_1,...,\psi _s+\kappa _s\bar
{\theta}\theta =c_s$ in ${\bf C}^{m|1}$). We can conclude, that $A$-model
having a smooth hypersurface in toric manifold as a target space is
equivalent to an $A$-model with toric supermanifold as a target space.
Similar statement is true if a  hypersurface is replaced with a smooth
complete intersection in toric manifold.

    I am indebted to S.Elitzur, A.Givental, A.Giveon, E.Rabinovici and,
especially, to  M.Kontsevich for useful discussions.
   \vskip .1in
  \centerline{{\bf References}}
  \vskip .1in
  1. Witten, E.: Mirror manifolds and topological field theory, in:
Essays on mirror manifolds, ed. S.Yau, International Press (1992)

  2. Witten, E.: The N matrix model and gauged WZW models. Nucl. Phys.
B371, 191 (1992)

  3. Vafa, C., Witten, E.: A strong coupling test of S-duality, hep-th
9408074

  4. Kontsevich, M.:Enumeration of rational curves via torus action,
hep-th  940535

  5. Sethi, S.: Supermanifolds,rigid manifolds and mirror symmetry,
hep-th 9404186

  6. Audin, M.: The topology of torus actions on symplectic manifolds.
Birkh\"auser 1991

  7. Fulton, W.: Introduction to toric varieties. Princeton Univ. Press 1993

  8. Schwarz, A., Zaboronsky, O., in preparation

\end{document}